\documentclass[noshowpacs,amsmath,
twocolumn,
superscriptaddress,
8pt,
aps,prb]{revtex4-2}
\pdfoutput=1
\usepackage{setspace}
\usepackage{amsmath}
\usepackage{multirow}
\usepackage{graphicx}
\usepackage{upgreek}

\renewcommand{\figurename}{\textbf{Fig.}}
\usepackage{verbatim}
\usepackage{amsfonts}
\usepackage{amssymb}
\usepackage{epstopdf}
\usepackage{dcolumn}
\usepackage{xcolor}
\usepackage{verbatim}
\usepackage{siunitx}
\setcitestyle{super}
\DeclareGraphicsExtensions{.pdf,.eps,.png,.jpg,.mps}
\begin{document}
\title{Broadband microwave-rate dark pulse microcombs in dissipation-engineered LiNbO$_3$ microresonators}

\author{Xiaomin Lv$^{1,2\dagger}$, Binbin Nie$^{1\dagger}$, Chen Yang$^{1\dagger*}$, Rui Ma$^{3}$, Ze Wang$^{1}$, Yanwu Liu$^{1}$, Xing Jin$^{1}$,  Kaixuan Zhu$^{1}$, Zhenyu Chen$^{3}$, Du Qian$^{1}$, Guanyu Zhang$^{1}$, Guowei Lv$^{1}$, Qihuang Gong$^{1,4,5}$, Fang Bo$^{3,*}$, and Qi-Fan Yang$^{1,5,*}$\\
$^1$State Key Laboratory for Artificial Microstructure and Mesoscopic Physics and Frontiers Science Center for Nano-optoelectronics, School of Physics, Peking University, Beijing, 100871, China\\
$^2$Hefei National Laboratory, Hefei, 230088, China\\
$^3$Nankai University, Tianjin, 300071, China\\
$^4$Collaborative Innovation Center of Extreme Optics, Shanxi University, Taiyuan, 030006,  China\\
$^5$Peking University Yangtze Delta Institute of Optoelectronics, Nantong, 226010, China\\
$^{*}$Corresponding author: ycoptics@pku.edu.cn; bofang@nankai.edu.cn; leonardoyoung@pku.edu.cn}

\date{}

\begin{abstract}
Kerr microcombs generated in optical microresonators provide broadband light sources bridging optical and microwave signals. Their translation to thin-film lithium niobate unlocks second-order nonlinear optical interfaces such as electro-optic modulation and frequency doubling for completing comb functionalities. However, the strong Raman response of LiNbO$_3$ has complicated the formation of Kerr microcombs. Until now, dark pulse microcombs, requiring a double balance between Kerr nonlinearity and normal group velocity dispersion as well as gain and loss, have remained elusive in LiNbO$_3$ microresonators. Here, by incorporating dissipation engineering, we demonstrate dark pulse microcombs with 25 GHz repetition frequency and 200 nm span in a high-$Q$ LiNbO$_3$ microresonator. Resonances near the Raman-active wavelengths are strongly damped by controlling phase-matching conditions of a specially designed pulley coupler. The coherence and tunability of the dark pulse microcombs are also investigated. Our work provides a solution to realize high-power microcombs operating at microwave rates on LiNbO$_3$ chips, promising new opportunities for the monolithic integration of applications spanning communication to microwave photonics.
\end{abstract}

\maketitle
Optical frequency combs, consisting of discrete spectral lines aligned to equidistant frequency grids, serve as a coherent link between optical and microwave signals \cite{diddams2020optical}. They have become the cornerstone for precision spectroscopy \cite{picque2019frequency} and metrology \cite{ludlow2015optical}, and their potential to enhance a diverse range of applications hinges on instrumental advancements that prioritize reduced size, weight, and power consumption. This aspiration has been realized in optical microresonators as microcombs, which harness resonantly-enhanced nonlinear optical processes for comb generation \cite{2011science}. To date, mode-locked microcombs have been demonstrated in many microresonator platforms, such as MgF$_2$ \cite{Herr2014Temporal}, silica \cite{yi2015soliton}, Si$_3$N$_4$ \cite{brasch2016photonic}, SiC \cite{guidry2022quantum}, and LiNbO$_3$ \cite{he2019self}. LiNbO$_3$ has become an appealing comb-generating material due to its various optical nonlinearities, broad transparency window, and compatibility with integrated photonics \cite{boes2023lithium}. The most well-known Pockels and Kerr effects have enabled electro-optic combs \cite{zhang2019broadband,rueda2019resonant,yu2022integrated} and soliton microcombs \cite{he2019self,gong2019soliton,gong2020photonic,gong2020near,lu2023two,song2024octave} in LiNbO$_3$ microresonators. Furthermore, their combination is expected to complete microcomb functionalities, enabling electro-optic tuning \cite{he2023high} and self-referencing that are critical for compact spectrometers \cite{suh2016microresonator,dutt2018chip,yang2019vernier,obrzud2019microphotonic,suh2019searching}, miniaturized optical clocks \cite{papp2014microresonator,newman2019architecture}, low-noise microwave generation \cite{kudelin2024photonic,sun2024integrated,jin2024microresonator}, and high-precision ranging \cite{suh2018soliton,trocha2018ultrafast,riemensberger2020massively}.

LiNbO$_3$ is a strong Raman-active crystal with several narrow-band vibrational energy levels (Extended Data Fig. \ref{figure7}). Continuously pumping the microresonator can trigger stimulated Raman scattering (SRS), potentially leading to Raman lasing in microresonators and thus complicating the comb formation \cite{Okawachi17,yu2020raman,gong2020photonic}. It is a prerequisite to mitigate SRS in LiNbO$_3$ microresonators for soliton microcomb generation \cite{gong2020photonic,he2023high}, and the situation becomes more challenging for another important class of Kerr microcombs known as dark pulse microcombs. Unlike the sech$^2$-shaped soliton microcombs, dark pulse microcombs appear as flat-top pulses in the microresonator \cite{xue2015mode,lobanov2015frequency,kim2019turn,helgason2021dissipative,jin2021hertz,lihachev2022platicon}, exhibiting superior pump-to-comb conversion efficiency and high comb line power. Although these features are appealing for communications \cite{fulop2018high,jorgensen2022petabit,rizzo2023massively} and microwave photonics \cite{shu2022microcomb}, the comparatively higher intracavity power of dark pulse microcombs also promotes SRS, especially in microresonators with small free spectral ranges (FSRs) where optical modes more easily resonate with Stokes photons \cite{Okawachi17,zhao2023widely}. To date, dark pulse microcombs remain challenging to realize in LiNbO$_3$ microresonators.

In this article, we present a dissipation engineering approach to mitigate SRS in an integrated high-Q LiNbO$_3$ microresonator, where pulley waveguide couplers are designed to introduce wavelength-dependent losses to the microresonator. Consequently, we demonstrate the generation of dark pulse microcombs with repetition frequencies as low as 24.8 GHz and spectra spanning over 200 nm. Furthermore, we show the tuning of repetition frequency over 107 kHz by adjusting the laser-cavity detuning and characterize the phase noise performance, highlighting their potential for various applications at optical and microwave frequencies.


\begin{figure*}[!ht]
\centering
\setlength{\abovecaptionskip}{0 cm}
    \includegraphics[width=\linewidth]{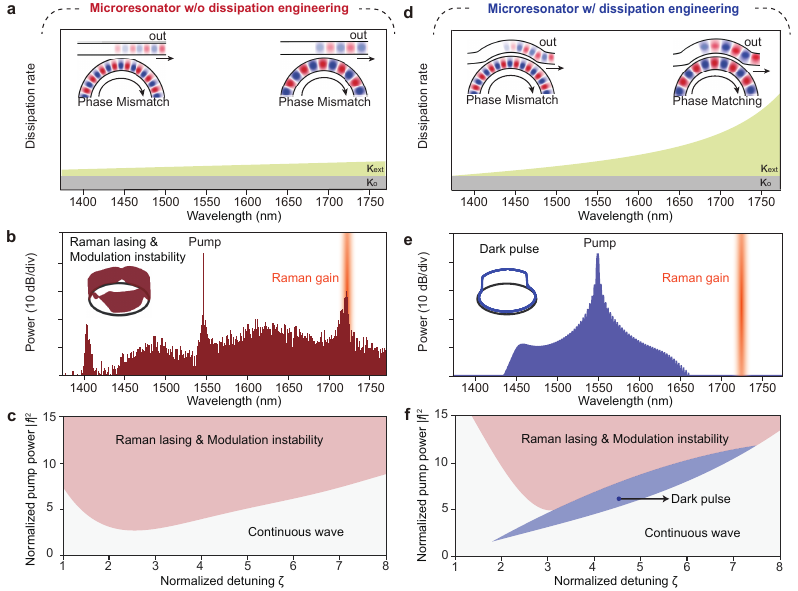}
    \caption{{\bf Dissipation engineering for dark pulse microcomb generation.} {\bf a,} Dissipation rate versus wavelength for microrings coupled to straight bus waveguides. $\kappa_o$: intrinsic loss; $\kappa_\mathrm{ext}$: external coupling loss. {\bf b,} Simulated optical spectrum and temporal profile of microcomb in microresonators without dissipation engineering. The color gradient indicates the Raman gain. {\bf c,} Simulated phase diagram as a function of pump power and laser-cavity detuning in microresonators without dissipation engineering. The pump power and the detuning are normalized to the parametric oscillation threshold and half the linewidth of the pump mode, respectively. The involved parameters are further defined in Methods. {\bf d,} Dissipation rate versus wavelength for microrings coupled to pulley bus waveguides. {\bf e,} Simulated optical spectrum and temporal profile of microcomb in microresonators with dissipation engineering. {\bf f,} Simulated phase diagram as a function of pump power and laser-cavity detuning in microresonators with dissipation engineering.}  
    \label{figure1}
\end{figure*}
\begin{figure*}[!ht]
\centering
   \includegraphics[width=\textwidth]{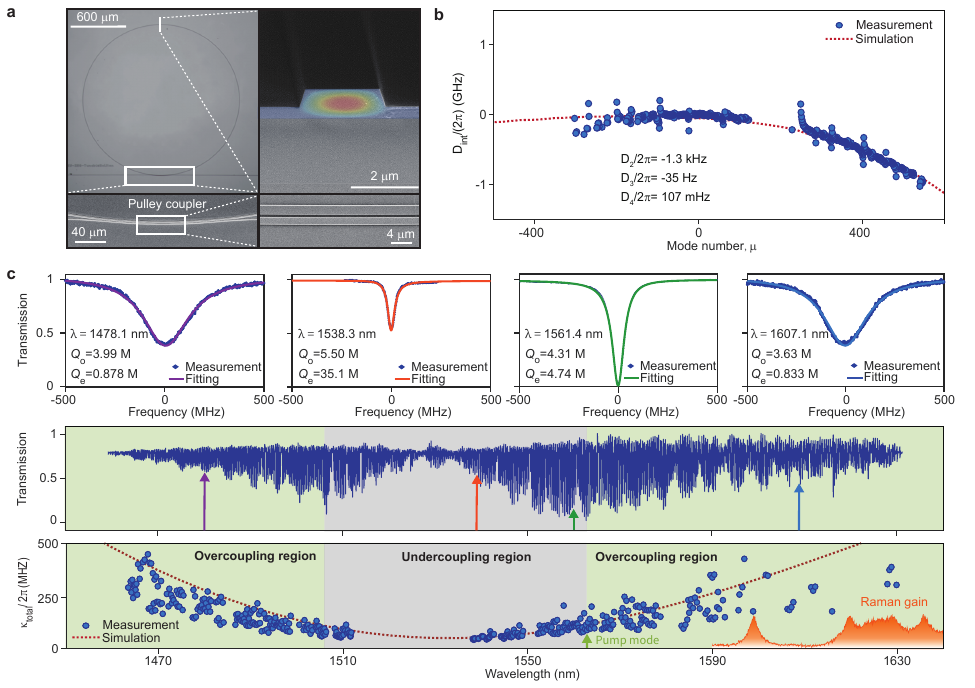}
   \caption{{\bf Device design and characterization.} 
   {{\bf a,} Optical microscope images of LiNbO$_3$ microring (left, top) and scanning electron microscope images detailing various components: waveguide-coupled region (left, bottom), the cross-section of the microring waveguide (right, top), and a zoom-in view of the pulley waveguide coupler (right, bottom). The intensity profile of the fundamental TE mode is overlaid on the cross-section. {\bf b,} Integrated dispersion ($D_\mathrm{int}$) of the fundamental TE mode family. The values of $D_{2}$, $D_{3}$, and $D_{4}$ are also listed. {\bf c,} Top panel: normalized transmission spectra of four typical resonances at different wavelengths. Middle panel: broadband transmission spectrum. Bottom panel: measured and simulated total dissipation rates as a function of wavelength. Color shadings represent different coupling conditions. The pump mode and the corresponding Raman gain are also indicated.}}
   \label{figure2}
\end{figure*}

\noindent
\begin{figure*}[hbtp]
\centering
    \includegraphics[width=\linewidth]{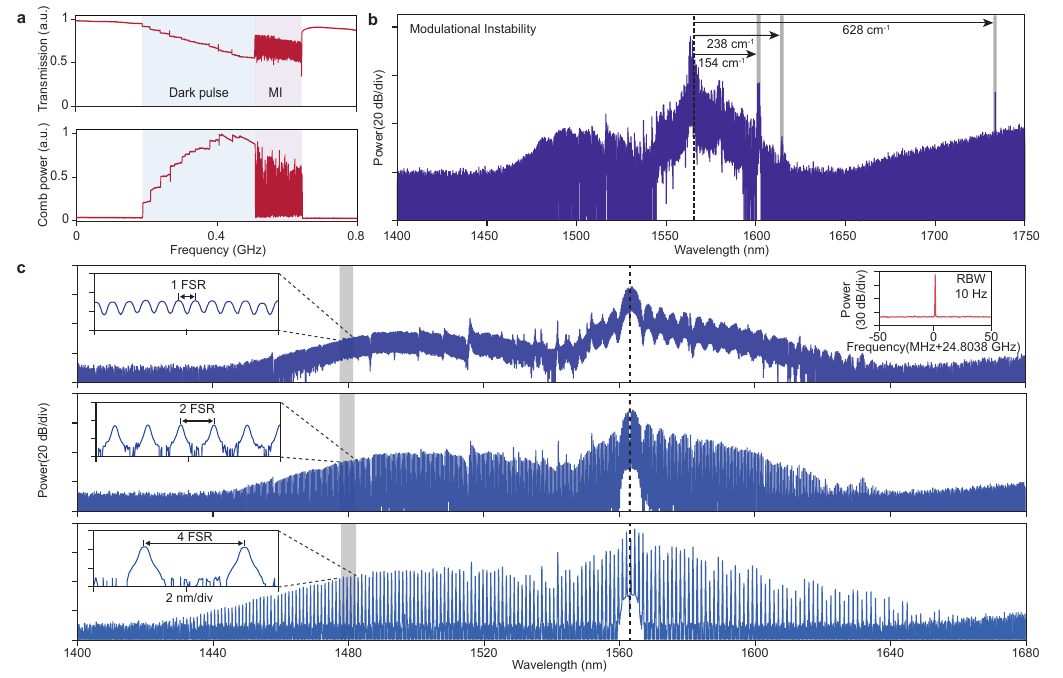}
    \caption{{\bf Formation of dark pulse microcombs.} 
    {\bf a,} Normalized transmitted pump power (top panel) and comb power (bottom panel) as a function of the increased pump laser frequency. {\bf b,} Optical spectrum of microcomb at MI state. Three Raman lasing modes are labeled with corresponding Raman shifts of 154 cm$^{-1}$, 238 cm$^{-1}$, and 628 cm$^{-1}$. {\bf c,} Optical spectra of dark pulse microcombs with different comb line spacing. The left insets show the zoom-in views of the comb lines. The right inset shows the measured electrical beatnote of the 1-FSR microcomb with 10-Hz resolution bandwidth (RBW).}
    \label{figure3}
\end{figure*}

\begin{figure*}[!ht]
    \includegraphics[width=\textwidth]{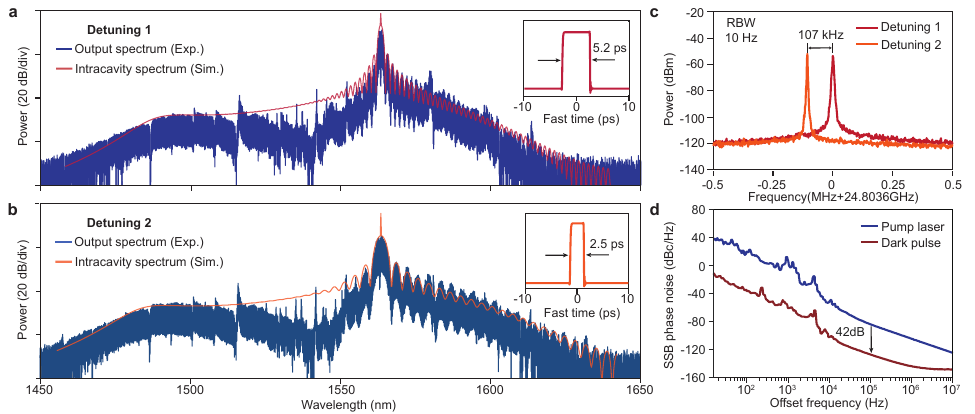}
    \caption{{\bf Tuning and coherence of dark pulse microcombs.} {\bf a,b,} Optical spectra of dark pulse microcombs for two different laser-cavity detunings. The simulated spectral envelopes of the microcombs are also plotted. The detunings are inferred from the simulations as 196 MHz and 231 MHz for ({\bf a}) and ({\bf b}), respectively. Insets: simulated intracavity waveforms. {\bf c,} Electrical beatnotes of dark pulse microcombs. The RBW is 10 Hz. {\bf d,} Typical phase noise of the repetition rate of the dark pulse microcomb and the pump laser.}
    \label{figure4}
\end{figure*}

\vspace{6pt}
{\bf\noindent Results}

{\bf\noindent Dissipation engineering}

\noindent The optical dissipation in microresonators comprises both intrinsic and external components. The former ($\kappa_\mathrm{o}$) is primarily influenced by material absorption and surface scattering, which is challenging to engineer in the frequency domain. On the other hand, external dissipation ($\kappa_\mathrm{ext}$) is dictated by waveguide coupling, which can be tuned in the frequency domain by designing phase-matching conditions between the waveguide and microresonator. Therefore, controlling $\kappa_\mathrm{ext}$ provides a route for dissipation engineering. 

We first consider the most widely-used coupler design, i.e., straight waveguide couplers, whose coupling to a ring resonator results in a confined coupling region due to limited spatial overlap. As the wave propagates along the waveguide, the effective distance between the waveguide and microring rapidly increases, leading to phase mismatching and a gradual increase in $\kappa_\mathrm{ext}$ with wavelength due to modal leakage (Fig. \ref{figure1}a). We simulate comb formation in a normal-dispersion LiNbO$_3$ microresonator with a straight waveguide coupler based on the generalized Lugiato-Lefever equation augmented by SRS (see methods). The FSR of the microresonator is set significantly smaller than the Raman gain bandwidth. The nearly constant dissipation profile gives rise to Raman lasing when the microresonator is pumped by a continuous-wave laser (Fig. \ref{figure1}b). Additionally, four-wave mixing interactions between the pump and Raman laser initiate modulational instability (MI), resulting in chaotic waveforms within the microresonator. Numerical simulations using a broader selection of pumping parameters (pump power and laser-cavity detuning) reveal the phase diagram, which excludes dark pulse microcombs (Fig. \ref{figure1}c).

This situation can be overturned by employing pulley waveguide couplers, which wrap around the resonator with a consistent gap, thereby considerably enhancing the coupling length compared to traditional straight waveguide coupling (Fig. \ref{figure1}d). By carefully designing the geometry of the waveguide and the microring, it is possible to achieve phase-matching conditions at specific wavelengths. Consequently, $\kappa_\mathrm{ext}$ experiences a rapid increase towards the phase-matching wavelength. When modes within the Raman gain bandwidth experience much larger dissipation rates than modes near the pump, dark pulse microcombs can form in the microresonator, with a typical optical spectrum shown in Fig. \ref{figure1}e. Despite the presence of Raman lasing and modulational instability at high pump power levels, a distinct region indicating the existence of dark pulse microcombs emerges in the phase diagram (Fig. \ref{figure1}f), highlighting the efficacy of dissipation engineering.

\vspace{6pt}
{\bf\noindent Device characterization}

\noindent We fabricate dissipation-engineered microring resonators on 570-nm-thick z-cut lithium niobate on insulator (LNOI) wafers (Fig. \ref{figure2}a). The geometry of the microresonator, characterized by a radius of 817 \textmu m, an etching depth of 300 nm, and a top width of 2.2 \textmu m, is chosen to induce normal group velocity dispersion (GVD). Experimentally, we measure the resonant frequencies of the fundamental transverse-electric (TE) mode family over a wavelength range of 1460 nm to 1630 nm using tunable continuous-wave lasers calibrated by a Mach-Zehnder interferometer. The measured integrated dispersion, defined as $D_\mathrm{int}=\omega_\mu-\omega_0-D_1 \mu=\sum_{n\geq2} D_n\mu^n/n!$, where $\mu$ is mode index relative to the pump mode, $\omega_\mu$ represents the resonant frequency of the $\mu_\mathrm{th}$ mode, and $D_1/2\pi$ is the free spectral range (FSR), is plotted in Fig. \ref{figure2}b. The FSR of the microresonator is found to be 24.803 GHz. Polynomial fitting of the integrated dispersion data reveals the second-order dispersion $D_2/2\pi=-1.3$ kHz, third-order dispersion $D_3/2\pi=-35$ Hz and fourth-order dispersion $D_4/2\pi=107$ mHz. These results align well with our numerical simulations.

The pulley coupler, characterized by a bus wavegude width of 1.756 µm, a pulley angle of 15 degrees, and a constant gap of 450 nm (Extended Data Fig. \ref{figure5}), is anticipated to exhibit increased coupling losses at wavelengths exceeding 1600 nm. These wavelengths feature strong Raman gain for 1563.2 nm pump lasers, as depicted in Fig. \ref{figure2}c and Extended Data Fig. \ref{figure7}. To investigate the impact of the pulley waveguide, we analyze the broadband transmission spectrum of the microresonator. Given that the intrinsic loss remains relatively stable within this spectral range, we utilized this knowledge ($Q_o\approx4\sim5$ M) to distinguish between intrinsic and external coupling losses by applying Lorentzian fitting to the transmission spectra. Our analysis reveals that the microresonator transitions from over-coupling to under-coupling and then returns to over-coupling as the wavelength increases from 1460 nm to 1630 nm. Notably, the total dissipation rates reach $2\pi\times350$ MHz at 1600 nm, representing a threefold increase compared to those observed near 1563.2 nm. The excellent agreement between our measured and simulated results underscores the precision with which the device geometry is controlled during fabrication.

\vspace{6pt}
{\bf\noindent Generation of dark pulse microcombs}

\noindent An amplified continuous-wave laser pumps the microresonator at around 1563.2 nm for comb generation. The laser is scanned from the red-detuned side towards the blue-detuned side of the mode. The transmission spectra of the pump and other comb lines are selected using a tunable fiber Bragg-grating filter (Fig. \ref{figure3}a). Low-noise steps appear first on the transmission spectra, corresponding to mode-locked microcombs. Further increasing the laser frequency results in chaotic transmission spectra, referred to as MI. As shown in Fig. \ref{figure3}a, the optical spectra of MI manifest three Raman lasers with frequency shifts of 154 cm$^{-1}$, 238 cm$^{-1}$, and 628 cm$^{-1}$, respectively, corresponding to the optical phonon modes of E(TO)$_{1}$, E(TO)$_{2}$, and A(TO)$_{4}$ (Extended Data Fig. \ref{figure7}b) \cite{10.1063/1.1652565,schaufele1966raman,barker1967dielectric}.  Other sidebands arise from the four-wave mixing processes and are not coherent with each other.

The photorefractive effect in LiNbO$_3$ allows for bi-directional accessing the low-noise steps in terms of tuning the laser frequency \cite{he2019self}. Each step corresponds to a different state of the dark pulse microcomb, and several representative optical spectra are displayed in Fig. \ref{figure3}c. The combs span from 1440 nm to 1640 nm, and Raman lasing actions are not observed. The spacings of the microcombs can be adjusted from single FSR to 4 FSR. Microcombs with multiple-FSR spacing exhibit enhanced power per comb line and an extended spectral span captured by the optical spectral analyzer \cite{lu2021synthesized}. A few spikes are present on the spectra due to the coupling between the fundamental TE mode and other transverse modes \cite{yang2016spatial}. Photodetection of the single-FSR microcomb reveals a monotone electrical beatnote at 24.8038 GHz, corresponding to the repetition frequency of the microcomb. It unambiguously demonstrates the mode-locking of the microcomb.

\vspace{6pt}
{\bf\noindent Tunability and coherence}

\noindent We fine-tune the microcomb with single-FSR spacing by adjusting the laser-cavity detuning. The spectra acquired at two different detunings exhibit characteristic modulations with different periods determined by the duration of the dark pulse \cite{xue2015mode,lao2023quantum} (Fig. \ref{figure4}a-b). To infer the temporal profiles of the two states, we perform numerical simulations based on experimentally measured device parameters. The fittings of the optical spectra reveal flat-top pulses in the microresonator, with duration varied from 5.2 ps to 2.5 ps depending on the detuning. The deviation between the simulated intracavity spectra and the experimental results is attributed to the wavelength-dependent coupling to the bus waveguide. For instance, the dips observed in the experimental spectra near 1528 nm aligned with the under-coupled regime shown in Fig. \ref{figure2}c. We further characterize the microwave performance of the two states. Their electrical beatnotes are found to differ by 107 kHz (Fig. \ref{figure4}c). Such tuning is due to a combination of higher-order dispersion and photorefractive effects. Using a phase noise analyzer, we measure the phase noise of the beatnote, as plotted in Fig. \ref{figure4}d. The phase noise reaches -102 dBc/Hz at 10 kHz offset frequency and goes down to -143 dBc/Hz at 1 MHz offset frequency. The phase noise of the repetition rate is mostly derived from that of the pump laser, with a transduction factor around -42 dB. 

\vspace{6pt}
\noindent\textbf{Discussion}

\noindent The demonstration of dark pulse microcombs in LiNbO$_3$ microresonators has several benefits. Firstly, it alleviates the need for dispersion engineering to generate Kerr microcombs. Currently, realizing anomalous GVD in LiNbO$_3$ microresonators at visible wavelengths remains challenging. Dark pulse microcombs are compatible with the entire transparency window of LiNbO$_3$, which can facilitate the development of compact optical clocks \cite{ludlow2015optical}, astronomical spectroscopy \cite{obrzud2019microphotonic,suh2019searching}, and other interdisciplinary research \cite{eid2020dental}. Besides, these dark pulse microcombs can be potentially integrated with high-speed LiNbO$_3$ modulators \cite{wang2018integrated}. The resulting LiNbO$_3$-based transmitters offer numerous high-power comb lines and a decent signal-to-noise ratio for massively parallel data transmission \cite{fulop2018high,jorgensen2022petabit,rizzo2023massively}. Furthermore, while this work utilizes the Kerr effect, it is anticipated that the dark pulse microcomb can be augmented using the Pockels effect in LiNbO$_3$. This approach promises high-speed electro-optic tuning and modulation of the repetition frequency, which has been demonstrated in bright soliton microcombs \cite{he2023high}. It also provides an intriguing platform for exploring the interplay between the Kerr and Pockels effects in the comb-forming regime. \cite{hu2022high,englebert2023bloch}.

\bigskip

\noindent\textbf{Methods}

\begin{footnotesize}

\noindent\textbf{Raman spectroscopy.} The Raman spectra of LiNbO$_3$ are measured using Raman microscopes (LabRAM HR Evolution, Horiba). The conditions are recorded using Porto notation composed of four elements such as X(YY)X. The first X denotes the direction of incident laser propagation, the first Y represents the polarization direction of the incident laser, the second Y indicates the polarization of scattered Raman light, and the second X specifies the propagation direction of the scattered Raman light. During the measurement, the polarization of the incident laser is fixed perpendicular to the z-axis, and the analyzer is set to transmit y-polarized light for X(YY)X configuration and z-axis polarized light for X(YZ)X scattering (Extended Data Fig. \ref{figure7}). Since the z-axis is the optical axis, X(YY)X configuration corresponds to Raman scattering from TE modes to TE modes in z-cut microresonators. Likewise, X(YZ)X configuration corresponds to Raman scattering from TE modes to TM modes in z-cut microresonators.

\vspace{6pt}
    
\noindent\textbf{Device design.} Extended Data Fig. \ref{figure5} outlines the overall design process. First, we aim to achieve normal dispersion by carefully optimizing the width and etch depth of the microring. Subsequently, the width and gap of the pulley puller are adjusted to induce phase-matching at a specific wavelength of 2140 nm. Finally, the coupling length is fine-tuned to ensure appropriate coupling losses, such as critical coupling at pump wavelengths. All simulations were performed using finite-element methods.

\vspace{6pt}

\noindent\textbf{Device fabrication.} Devices are fabricated from a commercial Z-cut LiNbO$_3$-on-insulator wafer (NANOLN) with a 570-nm-thick LiNbO$_3$ layer on top of a 4.7-$\mu$m-thick silica buffer layer. Electron-beam lithography is used to define the patterns of bus waveguides and microrings in ma-N 2405 resist with a thickness of 750 nm. The resist pattern is transferred to the LiNbO$_3$ film using Ar$^+$ plasma with an etching depth of 300 nm. The etching process yields a characteristic 60$^\circ$ angle of the sidewalls. The photoresist is removed by organic reagent and oxygen plasma. Finally, the redeposition on the surface of the sample is removed using ammonia solution.

\vspace{6pt}

\noindent\textbf{Numerical simulations.} The generalized Lugiato-Lefever equation (g-LLE), augmented with SRS and wavelength-dependent losses, provides a framework for simulating the evolution of intracavity optical fields \cite{Herr2014Temporal,agrawal2007nonlinear,lugiato1987spatial,lin2007nonlinear}. This equation is expressed as
\begin{equation}
\begin{aligned}
    \frac{\partial A(\phi, t)}{\partial t} = &-i \delta \omega A - \mathcal{F}^{-1}\left\{\frac{\kappa(\mu)}{2} \Tilde{A}_\mu \right\} + \sqrt{\frac{\kappa_{\mathrm{ext}}(0) P_{\mathrm{in}}}{\hbar \omega_0}}\\
    &+ i \frac{D_2}{2} \frac{\partial^2 A}{\partial \phi^2} + i(1-f_R) g_o |A|^2 A \\
    &+ i g_o \frac{f_R A}{D_1} \int h_R\left(\frac{\phi-\phi^{\prime}}{D_1}\right) |A(\phi^{\prime})|^2 d \phi^{\prime}.
\end{aligned}
\end{equation}
The slowly-varying field envelope $A(\phi, t)$ is normalized to photon number and studied in the rotation frame which rotates around the microresonator at the rate $D_1$. $\Tilde{A}_\mu$ is the optical field of the $\mu_\mathrm{th}$, which is defined as $A(\phi, t)=\sum_\mu \Tilde{A}_\mu e^{i \mu \phi}$. $\delta \omega$ is the frequency detuning of the pump laser relative to the mode. The pump power is denoted by $P_{\text{in}}$. The Kerr nonlinear coefficient is defined as $g_o = {\hbar \omega_0^2 n_2 D_1}/{2\pi n_o A_{\text{eff}}}$, where $n_o$ and $n_2$ are the linear and nonlinear refractive indices, respectively. $A_\mathrm{eff}$ is the nonlinear effective mode area. The total optical dissipation of the $\mu_{\text{th}}$ mode, $\kappa(\mu) = \kappa_o(\mu) + \kappa_{\text{ext}}(\mu)$, accounts for both intrinsic ($\kappa_o(\mu)$) and external coupling ($\kappa_{\text{ext}}(\mu)$) losses. The wavelength-dependent loss term is implemented in the frequency domain for convenience, with $\mathcal{F}^{-1}\{ \cdot \}$ representing the inverse Fourier transform.

LiNbO$_3$ has several vibrational modes as shown in Extended Data Fig. \ref{figure7}. For simplicity but without loss of generality, in the simulation, we only include the mode with the strongest Raman response, A(TO)$_4$. The fractional contribution of Raman response to the total nonlinear polarization is denoted by $f_R$. The Raman response is formulated as a convolution product of the Raman response function and the temporal intracavity optical intensity. A standard form of the Raman response function for a single vibrational mode, $h_R(t)$, is given by \cite{agrawal2007nonlinear,lin2007nonlinear}
\begin{equation}
h_R(t) = \frac{\tau_1^2 + \tau_2^2}{\tau_1 \tau_2^2} \exp\left(-\frac{t}{\tau_2}\right) \sin\left(\frac{t}{\tau_1}\right)
\end{equation}

Numerical simulation of the g-LLE is performed using the split-step Fourier method, which involves 4096 modes. The parameters used to simulate the phase diagram in Fig. \ref{figure1} are: $\kappa(\mu)/2\pi = (0.0016\mu^2-0.5011\mu+106.0)$ MHz, $\omega_0/2\pi=191.8$ THz, $D_1/2\pi=25$ GHz, $D_2/2\pi=-5$ kHz, $n_o=2.0$, $n_2=1.8\times10^{-19} \mathrm{m^2/W}$, $A_\mathrm{eff}=1.135\ \mathrm{{\mu m}^2}$, $f_R=0.05$, $\tau_1=8.3\ \mathrm{fs}$, $\tau_2=544\ \mathrm{fs}$ \cite{barker1967dielectric,schaufele1966raman,kaminow1967quantitative,bache2012review}. In Fig. \ref{figure1}, the normalized detuning is defined as $\zeta={2 \delta \omega}/{\kappa(0)}$ and the normalized pump power is defined as $|f|^2=8g_o \kappa_\mathrm{ext}(0) P_\mathrm{in}/ \kappa^3(0) \hbar \omega_0$. In Fig. \ref{figure4}a, the detuning is set $\delta \omega_1/2\pi = 196$ MHz. In Fig. \ref{figure4}b, the detuning is set $\delta \omega_2/2\pi = 231$ MHz. Other parameters used in the simulation of Fig.
\ref{figure4} are the same as those used in simulating the phase diagram of Fig. \ref{figure1}.

\end{footnotesize}
\medskip

\bibliography{citeref}

\medskip
\noindent\textbf{Data availability}

\begin{footnotesize}
\noindent The data that support the plots within this paper and other findings of this study are available from the corresponding author upon reasonable request.
\end{footnotesize}

\medskip

\noindent\textbf{Code availability}

\begin{footnotesize}
\noindent The codes that support the findings of this study are available from the corresponding authors upon reasonable request. 
\end{footnotesize}

\medskip

\noindent\textbf{Acknowledgments}

\begin{footnotesize}
\noindent 
The authors thank Xinrui Luo, Zhenyu Xie, and Zhe Lv for helpful discussions. This work was supported by National Key R\&D Plan of China (Grant No. 2021YFB2800601), National Natural Science Foundation of China (92150108, 12304412), Beijing Natural Science Foundation (Z210004), the high-performance computing Platform of Peking University, and the Advanced Photonic Integrated Center (APIC) of State Key Laboratory of Advanced Optical Communication System and Networks.
\end{footnotesize}

\medskip

\noindent\textbf{Author contributions} 

\begin{footnotesize}
\noindent The experiment was conceived by X.L., B.N., and Q.-F.Y. Measurements were performed by X.L. and B.N. with assistance from Z.W. and K.Z. The Raman spectroscopy was performed by B.N. and C.Y. with assistance from G.Z. The device was fabricated by X.L, C.Y, and R.M. The device was designed by X.L. with assistance from Y.L. Numerical simulations of comb dynamics were performed by B.N. All authors contributed to writing the manuscript.
\end{footnotesize}

\vspace{6pt}
\noindent\textbf{Competing interests}

\begin{footnotesize}
\noindent The authors declare no competing interests.
\end{footnotesize}

\medskip

\noindent\textbf{Additional information}

\begin{footnotesize}
\noindent Supplementary information is available for this paper.

\medskip

\noindent {\bf Correspondence and requests for materials}  should be addressed to Q.-F.Y.

\newpage

\begin{figure*}[h!]
\setcounter{figure}{0}
\centering
    \includegraphics[width=\linewidth]{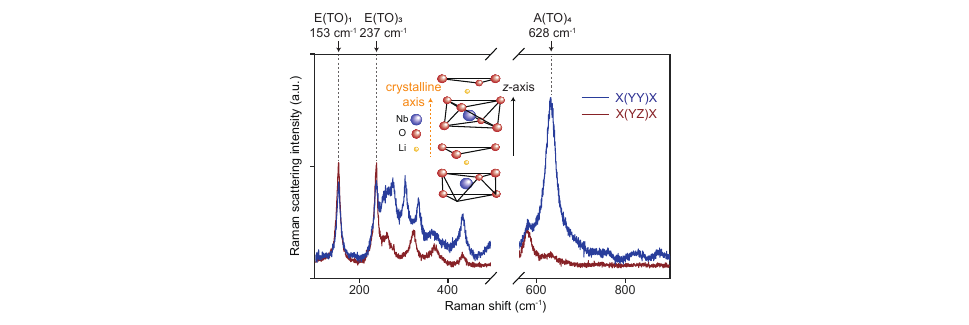}
    \renewcommand{\figurename}
    {\textbf{Extended Data Fig.}}
    \caption{{\bf Raman spectra of LiNbO$_3$.} {Raman scattering intensity is plotted as a function of Raman shift to depict the vibrational modes characterized by A$_1$ and E symmetries. The blue curve is obtained under the X(YY)X scattering configuration, while the red curve corresponds to the X(YZ)X scattering configuration. The Raman shift of 153 cm$^{-1}$, 237 cm$^{-1}$ and 628 cm$^{-1}$ correspond to the E(TO)$_{1}$ modes, the E(TO)$_{3}$ mode and the A(TO)$_{4}$ mode, respectively. Insets: crystal structure of LiNbO$_3$. The z-axis is the optical axis.}}
    \label{figure7}
\end{figure*}

\begin{figure*}[h!]
\centering
    \includegraphics[width=\linewidth]{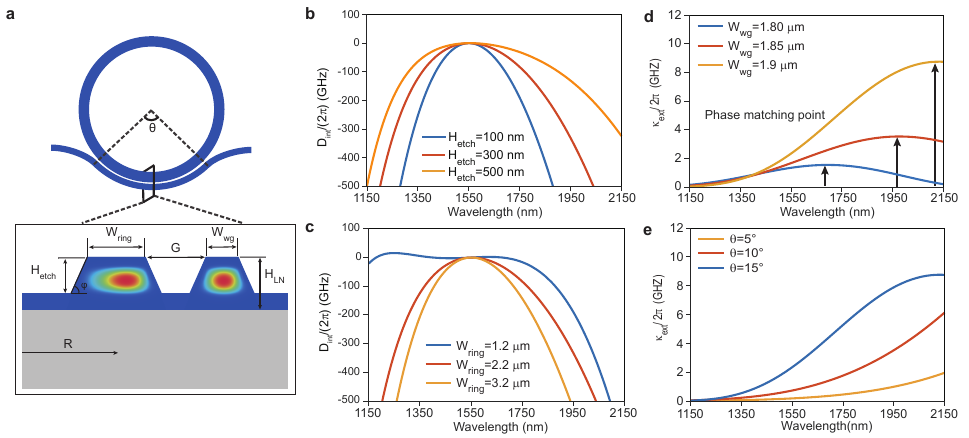}
    \renewcommand{\figurename}
    {\textbf{Extended Data Fig.}}
    \caption{{\bf Simulated dispersion and coupling characteristic of a pulley waveguide-coupled microring.} {\bf a,} Device geometry. W$_\mathrm{ring}$: the width of the ring resonator; W$_\mathrm{wg}$: the width of the waveguide; G: the gap between the waveguide and the ring resonator; H$_\mathrm{etch}$: the etching depth;  H$_\mathrm{LN}$: the thickness of the LiNbO$_3$ layer; R: the radius of the ring; $\theta$: the angle of the coupling region. The intensity profiles of the microresonator mode and bus waveguide mode are also overlaid. {\bf b, c,} Integrated dispersion D$_\mathrm{int}$ plotted versus wavelength for different H$_\mathrm{etch}$ ({\bf b}) and  W$_\mathrm{ring}$ ({\bf c}). {\bf d, e,} External dissipation rates $\kappa_\mathrm{ext}$ plotted versus wavelength for different W$_\mathrm{wg}$ ({\bf d}) and  $\theta$ ({\bf e}). The phase-matching wavelengths are indicated by arrows in {\bf d}.}
    \label{figure5}
\end{figure*}

\end{footnotesize}

\smallskip

\end{document}